\documentclass[twocolumn,showpacs,amsmath,amssymb,aps, prb,superscriptaddress]{revtex4-1}

\pdfoutput=1
\usepackage[]{graphicx}
\usepackage[]{verbatim}
\usepackage[]{color}
\usepackage{overpic}
\usepackage{rotating}
\usepackage{mathtools}
\usepackage{appendix}

\makeatletter
\newsavebox{\@brx}

\newcommand{\llangle}[1][]{\savebox{\@brx}{\(\m@th{#1\langle}\)}%
  \mathopen{\copy\@brx\mkern2mu\kern-0.8\wd\@brx\usebox{\@brx}}}
\newcommand{\rrangle}[1][]{\savebox{\@brx}{\(\m@th{#1\rangle}\)}%
  \mathclose{\copy\@brx\mkern2mu\kern-0.8\wd\@brx\usebox{\@brx}}}

  \newcommand{\lllangle}[1][]{\savebox{\@brx}{\(\m@th{#1\langle}\)}%
  \mathopen{\copy\@brx\copy\@brx\mkern4mu\kern-0.7\wd\@brx\usebox{\@brx}}}
\newcommand{\rrrangle}[1][]{\savebox{\@brx}{\(\m@th{#1\rangle}\)}%
  \mathclose{\copy\@brx\copy\@brx\mkern4mu\kern-0.7\wd\@brx\usebox{\@brx}}}

\graphicspath{{./}{./}}

\begin{document}
\title{Magnetic Orders Proximal to the Kitaev Limit in Frustrated Triangular Systems: Application to Ba$_3$IrTi$_2$O$_9$}
\author{Andrei Catuneanu}
\affiliation{Department of Physics and Center for Quantum Materials , University of Toronto, 60 St.~George St., Toronto, Ontario, M5S 1A7, Canada}
\author{Jeffrey G. Rau}
\affiliation{Department of Physics and Astronomy, University of Waterloo, Ontario, N2L 3G1, Canada}
\author{Heung-Sik Kim}
\affiliation{Department of Physics and Center for Quantum Materials , University of Toronto, 60 St.~George St., Toronto, Ontario, M5S 1A7, Canada}
\author{Hae-Young Kee}
\email{hykee@physics.utoronto.ca}
\affiliation{Department of Physics and Center for Quantum Materials , University of Toronto, 60 St.~George St., Toronto, Ontario, M5S 1A7, Canada}
\affiliation{Canadian Institute for Advanced Research, Toronto, Ontario, M5G 1Z8, Canada}

\begin{abstract}
Frustrated transition metal compounds in which spin-orbit coupling (SOC) and electron correlation work together have attracted much attention recently. In the case of 5$d$ transition metals, where SOC is large, $j_\text{eff}=1/2$ bands near the Fermi level are thought to encompass the essential physics of the material, potentially leading to a concrete realization of exotic magnetic phases such as the Kitaev spin liquid. Here we derive a spin model on a triangular lattice based on $j_\text{eff} = 1/2$ pseudo-spins that interact via antiferromagnetic Heisenberg ($J$) and Kitaev ($K$) exchanges, and crucially, an anisotropic $(\Gamma)$ exchange. Our classical analysis of the spin model reveals that, in addition to small regions of 120$^\circ$, $\mathbb{Z}_2$ / dual-$\mathbb{Z}_2$ vortex crystal and nematic phases, the stripy and ferromagnetic phases dominate the $J$-$K$-$\Gamma$ phase diagram. We apply our model to the 5$d$ transition metal compound, Ba$_3$IrTi$_2$O$_9$, in which the Ir$^{4+}$ ions form layered two-dimensional triangular lattices. We compute the band structure and nearest-neighbor hopping parameters using \textit{ab-initio} calculations. By combining our \textit{ab-initio} and classical analyses, we predict that Ba$_3$IrTi$_2$O$_9$ has a stripy ordered magnetic ground state.
\end{abstract}
\maketitle

\section{Introduction}
Transition-metal compounds in which electrons occupy $d$-orbitals have proven to be a vast playground for exotic and interesting physics. The non-trivial combination of lattice geometry, crystal field strength, spin-orbit coupling (SOC) and Coulomb repulsion between electrons in transition metals can conspire to produce novel ground states, including a plethora of magnetic orders\cite{witczak2014correlated, wan2011topological, Moon_Lab, TIstar, pesin2010mott, lawler2008SL, CK_arxiv, CLK_arxiv, KVCK2015}. Such materials, for instance the honeycomb family of iridates\cite{singh2010antiferromagnetic, longrangemoNIO2011, singh2012relevance, correlationsNIO2012, Choietal2012, comin_prl, clancyetal2012, gretarssonetal2013, gretarsson_prl2013}, have drawn attention \cite{shitade2009quantum, chaloupka2010kitaev, Kimchi2011, Bhattacharjee2012, mazin_prl, chkim_prl, Khaliullin_eg, foyevtsova2013ab} owing to the possibility of a concrete realization of Kitaev's exactly solvable spin-$1/2$ model \cite{kitaev2006anyons}. 

Recently, a general model of the layered honeycomb iridates was derived in which $j_{\rm eff} = 1/2$ states near the Fermi level captured the essential physics\cite{rau2014generic}. Arising from a combination of strong SOC and the octahedral crystal field environment for each Ir$^{4+}$ ion, these $j_{\rm eff} = 1/2$ states effectively behave as spin-$1/2$ pseudo-spins. It was found that a variety of magnetic ground states could be realized depending on the competition between Heisenberg ($J$), Kitaev ($K$) and symmetric off-diagonal $(\Gamma)$ spin exchanges at the nearest-neighbor level. The general principles applied to the honeycomb iridates can also be turned to triangular iridates which can harbor similar spin anisotropy.

Spins on a triangular lattice are inherently frustrated - potentially leading to new magnetic phases. It is therefore important to study the interaction between $j_{\rm eff} = 1/2$ pseudo-spins on a triangular lattice which have a microscopic origin similar to that in the honeycomb iridates. Classical and quantum studies of the Heisenberg-Kitaev (HK) model have been performed on the triangular lattice\cite{roushochatzakis2012, becker2015}, with possible applications to the triangular iridate Ba$_3$IrTi$_2$O$_9$\cite{Dey2012}. It has been shown that the HK model hosts an interesting $\mathbb{Z}_2$ vortex crystal phase which could be realized in Ba$_3$IrTi$_2$O$_9$; moreover, the presence of the Kitaev exchange destabilizes the 120$^\circ$ ordered phase of the Heisenberg model on the trianglular lattice. We will argue, however, that this model is incomplete and does not describe the physics of Ba$_3$IrTi$_2$O$_9$; a more comprehensive approach is required.

\begin{figure}[!ht]
  \includegraphics[width=0.45\textwidth]{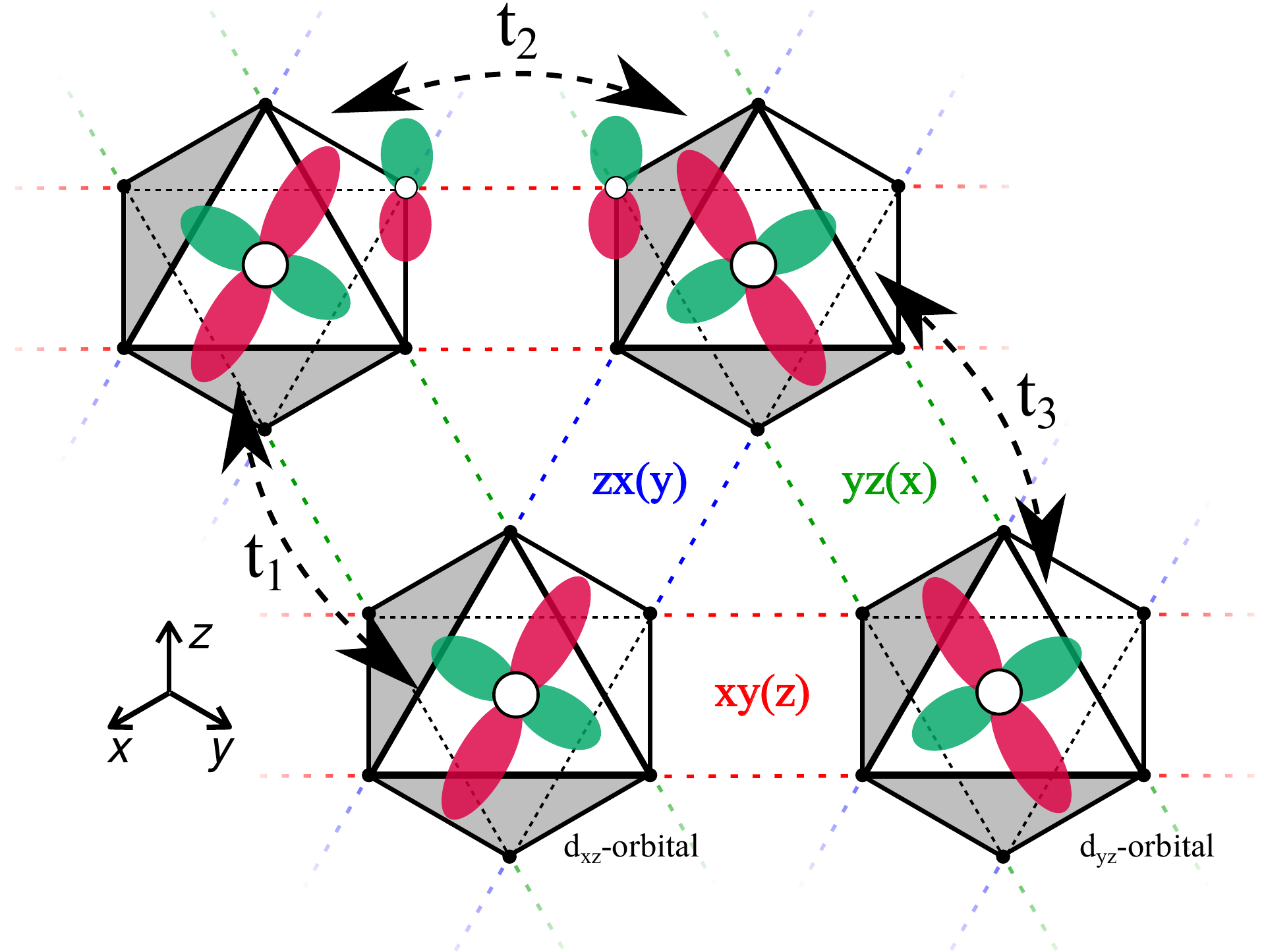}
  \caption{(Color online) (a) The major nearest-neighbor hopping channels in the limit of ideal octahedra. The parameters $t_1$ and $t_3$ describe two distinct $d$-orbital overlaps, while $t_2$ includes both $d$-orbital overlap and oxygen mediated hoppings. The green, blue and red dotted lines label the $x$, $y$ and $z$ bond types according to the convention $\alpha\beta(\gamma)$ in Eq. \eqref{JKG}.}
  \label{fig:hoppings}
\end{figure}

In this paper, we derive a spin model that is generic to layered triangular iridate compounds such as Ba$_3$IrTi$_2$O$_9$. Our derivation is founded upon the reasonable assumption that electron interaction strength ($U$) and Hund's coupling ($J_H$) provide the largest energy scales in the system, with SOC being the next largest energy scale, and electron hopping between $d$-orbitals as the smallest energy scale. In addition, we assume that the symmetries of an ideal triangular lattice of octahedra are respected in our derivation. From this foundation we derive a nearest-neighbor spin model with three spin exchanges: Heisenberg ($J$), Kitaev ($K$) and a symmetric off-diagonal exchange $(\Gamma)$. We map out the classical magnetic ground state phase diagram through a combination of Luttinger-Tisza\cite{LuttingerTisza} and classical Monte Carlo techniques and discuss its intricacies. 

We next apply our general model to the triangular iridate Ba$_3$IrTi$_2$O$_9$ by first performing \textit{ab-initio} calculations to determine its band structure with and without SOC. We then demonstrate that the states near the Fermi level are well described by $j_{\rm eff} = 1/2$ states, and we further estimate the nearest-neighbor tight-binding parameters. Consequently, we can estimate the region of the phase diagram that pertains to Ba$_3$IrTi$_2$O$_9$ (with and without the effects of octahedral distortion) and predict that it should exhibit stripy magnetic order in its ground state. Finally, experimental methods to test our prediction are also discussed. 

Our paper is organized in the following manner. We derive our general $j_{\rm eff}=1/2$ spin model on a triangular lattice and analyze its classical phase diagram in Section II. In Section III we discuss the crystal structure of Ba$_3$IrTi$_2$O$_9$, present the results of our \textit{ab-initio} calculations, estimate the strengths of the spin exchanges from \textit{ab-initio} tight-binding parameters, and predict the ground state for Ba$_3$IrTi$_2$O$_9$ using our model. In Section IV we examine the effects of octahedral distortion in Ba$_3$IrTi$_2$O$_9$ within our spin model. Finally, in Section V, we discuss and summarize the ramifications of these results and suggest possible experimental tests.

\section{Derivation of $j_{\rm eff}$=1/2 Spin Model on a Triangular Lattice \& Luttinger-Tisza and Classical Monte Carlo Analysis}

We begin by constructing a model of a triangular lattice of isolated Ir$^{4+}$ ions which are surrounded by ideal oxygen octahedra as in Fig. \ref{fig:hoppings}a. The local octahedral crystal field will split the Ir$^{4+}$ $5d$ levels into $t_{2g}$ and $e_{g}$ states. The large SOC of the Ir$^{4+}$ ions will further split the $t_{2g}$ manifold into completely filled $j_\text{eff} = 3/2$ states and half-filled $j_\text{eff} = 1/2$ states. Since the strong on-site Coulomb interactions localize these $j_\text{eff} = 1/2$ states, we can then consider the effective physics to be captured by interacting $j_\text{eff} = 1/2$ pseudo-spins. To arrive at this picture concretely, we start from a microscopic Hamiltonian which includes nearest-neighbor $d$-orbital hoppings as perturbations on top of an on-site Kanamori Hamiltonian\cite{kanamori1963electron} (see the Supplementary Material for more details). There are two types of $d$-orbital hopping at the nearest-neighbor level as shown in Fig. \ref{fig:hoppings}a: direct overlap between $d$-orbitals on Ir$^{4+}$ sites (denoted by $t_1$ and $t_3$), and oxygen mediated hopping via $p$-orbitals (denoted by $t_2$).

\begin{figure*}
  \centering
  \includegraphics[width=0.9\textwidth]{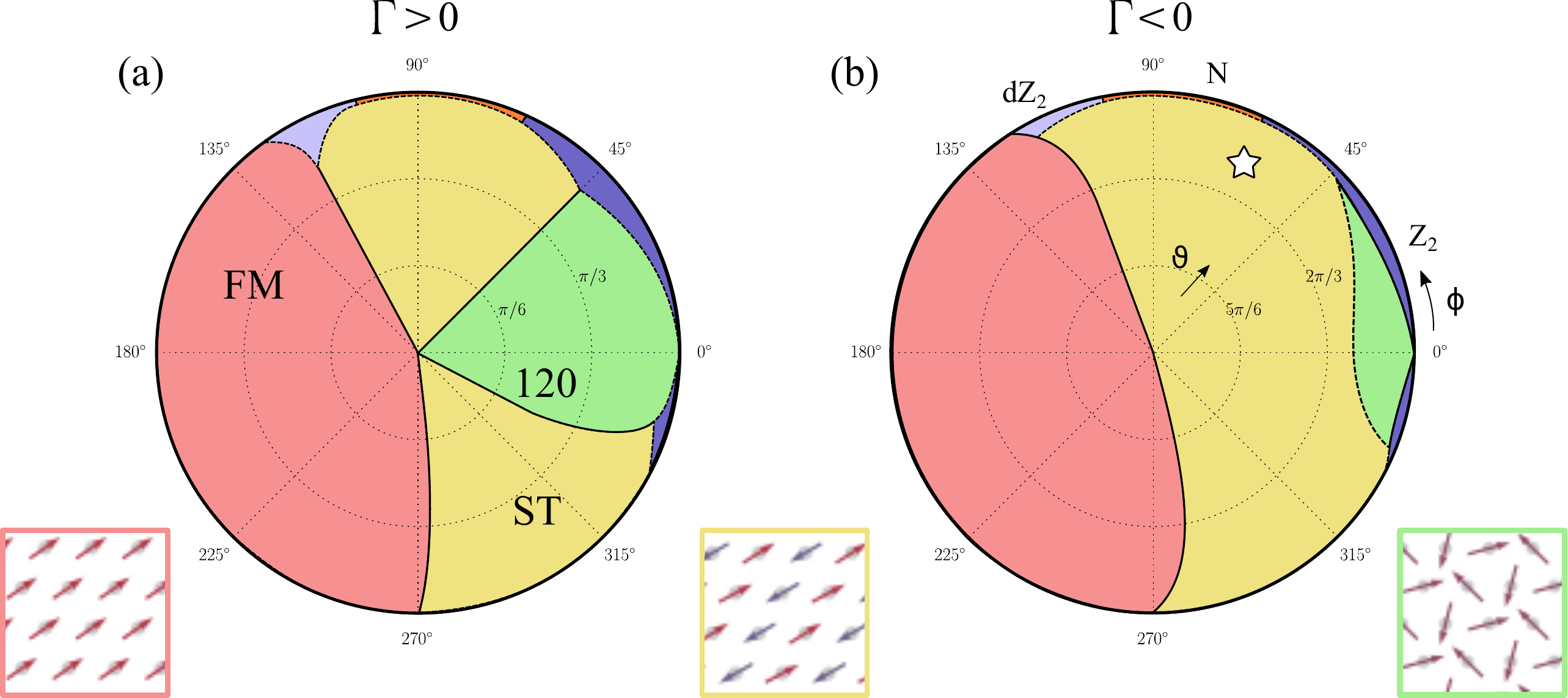}
  \caption{(Color online) Combined Luttinger-Tisza (LT) and Classical Monte Carlo (CMC) phase diagrams of the triangular lattice $J$-$K$-$\Gamma$ model for $\Gamma > 0$ (a) and $\Gamma < 0$ (b). The angles $\theta$ and $\phi$ denote the radial and azimuthal angles respectively. There are six phases represented: ferromagnet (FM), stripy (ST), 120 order (120), nematic (N), $\mathbb{Z}_2$ vortex ($Z_2$) and dual-$\mathbb{Z}_2$ ($dZ_2$). Real space spin representations of FM, ST and 120 orders are shown in the insets, while nematic,  $\mathbb{Z}_2$ vortex crystal and dual-$\mathbb{Z}_2$ vortex crystal phases are shown in the Supplementary Material. Phases bounded toward the edge ($\Gamma = 0$) by a dotted line are computed using CMC. The region marked by the white star is the relevant parameter regime for Ba$_3$IrTi$_2$O$_9$ predicted by our combined {\it ab-initio} and classical analysis.}
   \label{fig:phasediagrams}
\end{figure*}

The nearest-neighbor spin model takes the form
\begin{equation}
  H = \sum_{\alpha\beta(\gamma)\in\langle ij \rangle} \left[J {\bf S}_i\cdot {\bf S}_j + K S_i^\gamma S_j^\gamma +
  \Gamma (S_i^\alpha S_j^\beta + S_i^\beta S_j^\alpha)\right], \label{JKG}
\end{equation}
where $i$, $j$ denote Ir$^{4+}$ sites, and ${\bf S}_i$ is a $j_{\rm eff} = 1/2$ spin operator at a site with components $S_i^\alpha$ ($\alpha = x, y, z$). The exchanges $J$ and $K$ are Heisenberg and Kitaev exchanges respectively, while $\Gamma$ is a symmetric off-diagonal exchange (exact expressions can be found in the Appendix). We identify each bond as one of either $yz(x)$, $zx(y)$ or $xy(z)$ bonds, labeled by $\alpha\beta(\gamma)$. The spin components that interact on each bond depend on the bond type $\alpha\beta(\gamma)$.

We studied the phase space of the spin model using a combination of Luttinger-Tisza (LT) and classical Monte Carlo (CMC) analyses in order to determine the ground state for varying exchange parameters. We employed the simulated annealing technique in our CMC calculations, whereby the transition probability from state to state is determined by a temperature parameter that is slowly lowered until there is no improvement in the ground state energy. The classical approaches are considered simultaneously in Fig. \ref{fig:phasediagrams}. The LT and CMC analyses were found to agree with each other in both energy and spin configuration, except for small regions of the phase diagram computed by CMC (bounded by the dotted line toward the boundary), where LT could not identify the exact classical ground state. The exchanges are parameterized in spherical coordinates such that $J = \sin \theta \cos \phi, \ K = \sin \theta \sin \phi$ and $\Gamma = \cos \theta$ where $\theta \in [0,\pi]$, and $\phi \in [0,2\pi)$. The Heisenberg-Kitaev (HK) model is recovered in the limit in which $\theta = \pi/2$, where $\Gamma = 0$. We find six different magnetic orderings: ferromagnet (FM), stripy (ST), 120$^\circ$ order (120), nematic (N), $\mathbb{Z}_2$ vortex crystal and dual-$\mathbb{Z}_2$ vortex crystal. We depict the FM, ST, and 120 orders in the insets of Fig. \ref{fig:phasediagrams}, while nematic, $\mathbb{Z}_2$ vortex crystal and dual-$\mathbb{Z}_2$ vortex crystal phases are depicted in the Supplementary Material. A thorough description of $\mathbb{Z}_2$ and dual-$\mathbb{Z}_2$ orders can be found in Refs. \onlinecite{roushochatzakis2012, becker2015}. 
\begin{figure}[h]
  \includegraphics[width=0.45\textwidth]{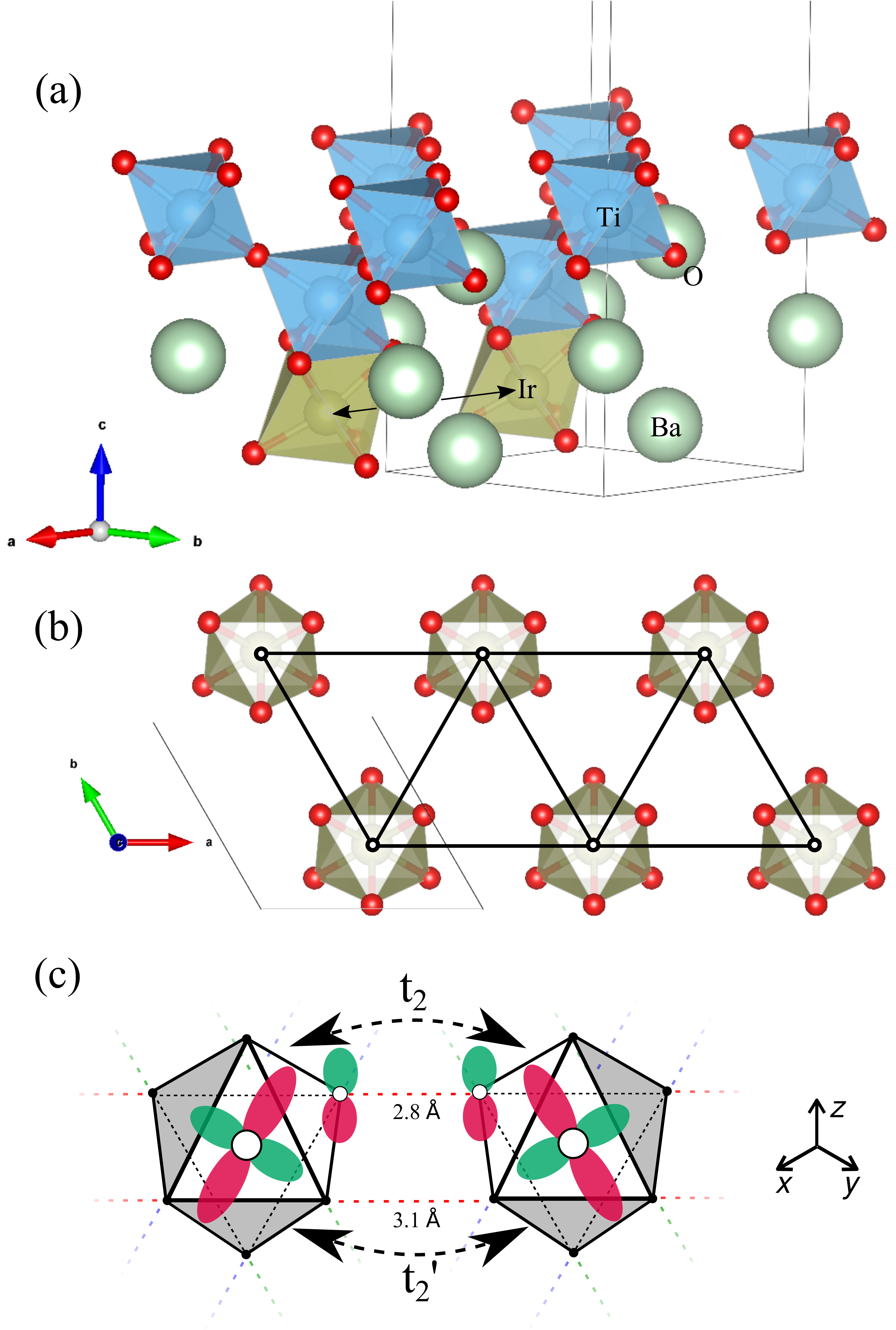}
  \caption{(Color online) (a) Half of the Ba$_3$IrTi$_2$O$_9$ primitive unit cell (bounded by the rectangular box) translated once in the a-direction featuring Ba (pale-green), Ti (blue), O (red) and Ir (yellow) atoms. There are two Ir$^{4+}$ ions in the unit cell, each belonging to different layered triangular lattices as in (b), with each site surrounded by oxygen octahedra. Nearest-neighbor Ir atoms belonging to different unit cells are connected by the black double-arrowed line. The Ir$^{4+}$ octahedra are face-shared with Ti$^{4+}$ octahedra. (c) Representation of distorted octahedra featuring two new hopping parameters, $t_2$ and $t_2'$, leading to a Dzyaloshinksii-Moriya (DM) exchange interaction in Eq. \eqref{distortedJKG}.}
  \label{fig:cell}
\end{figure}

  \begin{figure*}
  \centering
  \includegraphics[width=1.0\textwidth]{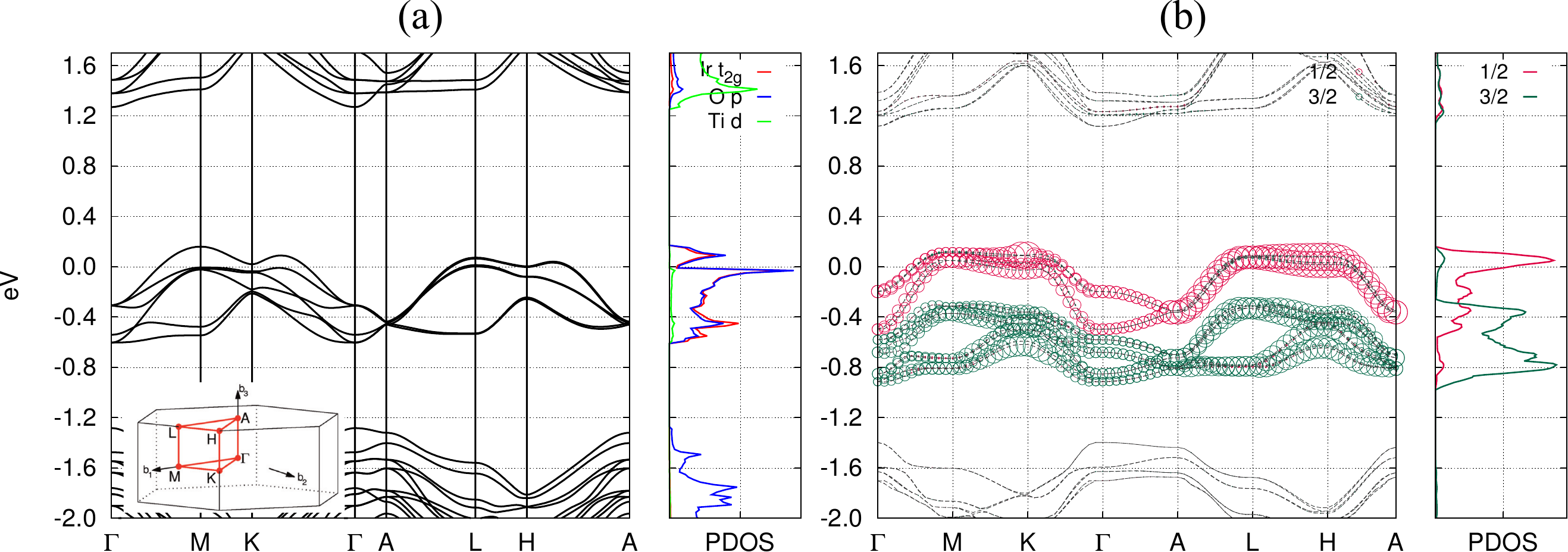}
  \caption{(Color online) Band structure calculations of Ba$_3$IrTi$_2$O$_9$ performed using OpenMX. (a) Band structure without SOC and projected density of states including Ir $t_{2g}$-, Ti d-, and O p-orbitals. The bands near the Fermi level are composed of contributions from Ir$^{4+}$ $t_{2g}$ orbitals and O p-orbitals. (b) Band structure with SOC and $j_\text{eff}$ projected density of states. The bands are projected onto $j_\text{eff}=1/2$ and $j_\text{eff}=3/2$ states using maximally-localized Wannier orbitals. The relative contribution of each state is represented by the intensity of the color. The $t_{2g}$ bands are well separated into $j_\text{eff}=1/2$ (pink) and $j_\text{eff}=3/2$ (teal) bands.}
   \label{fig:bandstructure}
\end{figure*}

  The boundaries of both phase diagrams in Fig. \ref{fig:phasediagrams} coincide, and show the classical ground states of the HK model ($\Gamma = 0$ in Eq. \eqref{JKG}). We first highlight some special points. The Heisenberg model ($J = 1$) hosts 120$^\circ$ order, agreeing with the known ground state on a triangular lattice. We also find that perturbing away from the Heisenberg model by adding $K$ tends to destabilize 120$^\circ$ order toward the previously discussed $\mathbb{Z}_2$ vortex crystal phase. The $K = 1$ point represents the Kitaev model on the triangular lattice. Here we find that the spins tend to order antiferromagnetically in chains along one of the three principal directions of the triangular lattice, with each chain decoupled from the rest. By flipping all the spins in any of the chains, the energy remains the same and so this state is highly degenerate - exponential in the size of the system. This classical ground state degeneracy has recently been analyzed and has been shown to be lifted by quantum fluctuations through an order-by-disorder mechanism, introducing a coupling between spins of next-nearest neighboring antiferromagnetic chains \cite{JackeliAvella2015}. Overall, the boundary of the phase diagram is in agreement with the previous Monte Carlo studies of the HK model on the triangular lattice. 

  However, when adding $\Gamma$, whether positive or negative, to the Heisenberg model, we find that 120$^\circ$ order emerges. For positive $\Gamma$, we find that 120$^\circ$ order persists until the degenerate $\Gamma = 1$ point, where 120$^\circ$, FM and ST orders meet. For the Kitaev model, on the other hand, the introduction of small $|\Gamma|$ immediately gives ST order as the magnetic ground state. The majority of the phase diagram upon the inclusion of $\Gamma$ is dominated by stripy or ferromagnet phases. 

\section{Application to B\lowercase{a}$_3$I\lowercase{r}T\lowercase{i}$_2$O$_9$}

Ba$_3$IrTi$_2$O$_9$ has a layered triangular lattice structure composed of Ir$^{4+}$ ions with a $d^5$ electron configuration surrounded by oxygen octahedra. It belongs to spacegroup $P6_3mc$, possessing a mirror plane and a screw axis along the c-direction, connecting the top and bottom halves of the unit cell. The bottom half of the unit cell is shown bounded by the rectangular box in Fig \ref{fig:cell}a. The IrO$_2$ layers are separated by two TiO$_2$ layers, with the Ir$^{4+}$ ions face-sharing their octahedra with the Ti$^{4+}$ octahedra directly above them along the c-axis. The Ba$^{2+}$ ions populate the unit cell among the IrO$_2$ and TiO$_2$ layers. There are two Ir$^{4+}$ ions belonging to the primitive unit cell, each of which belongs to different IrO$_2$ layers that form a triangular lattice, shown in Fig \ref{fig:cell}b. The distortion of the local octahedra, depicted in Fig \ref{fig:cell}c, is discussed in the next section. 

The local environment for the electrons at each site, with an octahedral crystal field, and large SOC, is reminiscent of that in Na$_2$IrO$_3$, Li$_2$IrO$_3$ and $\alpha$-RuCl$_3$. Therefore, we may expect that Ba$_3$IrTi$_2$O$_9$ can be described by similar physical principles involving pseudo-spin $j_\text{eff}=1/2$ states near the Fermi level. Unlike in honeycomb iridates, however, Ba$_3$IrTi$_2$O$_9$ forms triangular lattice layers and the oxygen octahedra surrounding the Ir$^{4+}$ ions are not edge-shared. Here we explain that the $j_\text{eff} = 1/2$ spin model derived in the previous section captures the essential physics of Ba$_3$IrTi$_2$O$_9$. To this end, we perform a series of band structure calculations for Ba$_3$IrTi$_2$O$_9$ and determine the nature of the electronic states near the Fermi level.

We used OpenMX\cite{openmx}, which implements the linear-combination-of-pseudo-atomic-orbitals method, to compute the band structure of Ba$_3$IrTi$_2$O$_9$ with and without SOC. A non-collinear DFT scheme and a fully relativistic $j$-dependent pseudopotential are used to treat SOC, with the Perdew-Zunger parameterization of the local density approximation (LDA) chosen for the exchange-correlation functional\cite{PerdewZunger1981}. We chose an energy cutoff of 300 Ry for the real-space sampling and used an $8\times 8\times 3$ $k$-grid for the Brillouin zone sampling. We used the maximally-localized Wannier orbitals method\cite{MarzariVanderbilt1997} implemented in OpenMX\cite{WengOzakiTerakura2009} to obtain the tight-binding Hamiltonian for Ir $t_{2g}$ orbitals.

The results of our calculation are shown in Fig \ref{fig:bandstructure}. The calculated band structure without SOC is shown in Fig \ref{fig:bandstructure}a alongside the projected density of states (PDOS). The bands near the Fermi level are composed of p-orbitals from oxygen and $t_{2g}$ orbitals coming from the Ir$^{4+}$ ions. A gap of approximately 1 eV separates these bands from the empty Ti $d$-bands. We note that the bandwidth is relatively small, reflecting the large distance between Ir$^{4+}$ sites.

To understand the effects of SOC, we also computed the band structure including SOC, shown in Fig \ref{fig:bandstructure}b with the bands and PDOS projected onto the $j_\text{eff}=1/2$ and $3/2$ states. We find that the $t_{2g}$ bands are split by SOC into well-separated $j_\text{eff}=1/2$ (pink) and $j_\text{eff}=3/2$ (teal) states. The five valence electrons will serve to completely fill the $j_\text{eff}=3/2$ states and half-fill the $j_\text{eff}=1/2$ states. We can therefore expect the effective physics of Ba$_3$IrTi$_2$O$_9$ to be well described by the pseudo-spin $j_\text{eff}=1/2$ model derived in the previous section.

Having established the $j_\text{eff}=1/2$ nature of Ba$_3$IrTi$_2$O$_9$, we can further estimate the hopping parameters coming from the major hopping channels depicted schematically in Fig \ref{fig:hoppings}a. We again use the maxially-localized Wannier orbital method to estimate the nearest-neighbor hopping parameters. In the limit of ideal local octahedra, we find that the nearest-neighbor hopping parameters are $t_1 = 7.4$ meV, $t_2 = -23.5$ meV and $t_3 = -119$ meV. All further neighbor hopping parameters are found to be small ($\lesssim 5$ meV), therefore, a tight binding model including only nearest-neighbor parameters will aptly describe the kinetic microscopics.

From our {\it ab-initio} calculations, we can estimate the strength of the exchanges knowing the strengths of the nearest-neighbor tight binding parameters. Taking on-site Coulomb repulsion to be $U = 2.0$ eV and fixing the ratio of Hund's coupling to $U$ as $J_H/U = 0.2$, we find $J \simeq 2$ meV, $K \simeq 4$ meV and $\Gamma \simeq -2$ meV for Ba$_3$IrTi$_2$O$_9$. Normalizing the exchange couplings by $N = \sqrt{J^2 + K^2 + \Gamma^2}$ (so that $J^2 + K^2 + \Gamma^2 = 1$), we equivalently find $J/N \simeq 0.36$, $K/N \simeq 0.86$ and $\Gamma/N = -0.36$, with the Kitaev exchange being the dominant exchange at the nearest-neighbor level. We also point out that $\Gamma$ is non-zero and negative. This area of phase space is marked by the white star in Fig. \ref{fig:phasediagrams}, and falls into the stripy region of magnetic order.

Our analysis has been performed under the most general considerations imposed by the symmetries of the lattice and the local electron environment at each site, arriving at the model in Eq. \eqref{JKG}. When $\Gamma = 0$, our model reduces to the HK model; however, the HK model is insufficient to describe Ba$_3$IrTi$_2$O$_9$ as seen in the significant $\Gamma$ that is of the same strength as the Heisenberg exchange $J$. Indeed, even a relatively small $\Gamma$ can act to unsettle the phases on the boundary of the phase diagram. Moreover, all layered triangular systems of this kind can be treated with our model and, barring fine-tuning of the tight binding parameters, $\Gamma$ is not zero in general. 

\section{Effects of Octahedral Distortion}
We also investigated the presence of the distortion in the oxygen octahedra, which breaks inversion symmetry about the Ir-Ir bond center. Such a distortion will induce two inequivalent oxygen mediated hoppings, $t_2$ and $t'_2$ shown in Fig. \ref{fig:cell}b, as well as other potential hoppings. From our {\it ab-initio} calculations, we estimate that $t_2 = -13$ meV and $t_2' = -32$ meV. In the case of Ba$_3$IrTi$_2$O$_9$, we only consider these hopping parameters since our {\it ab-initio} calculations suggest that other hopping channels created by the distortion have small amplitudes (less than half of $t_2$).

The main result of the non-ideal local crystal environment is to introduce a Dzyaloshinskii-Moriya (DM) exchange term to the spin model. It also breaks apart the Heisenberg-Kitaev exchanges into three anisotropic exchanges:
  \begin{align}
    H_\text{distorted} &= \sum_{\alpha\beta(\gamma) \in \langle ij\rangle} \Big[ J^\alpha S_i^\alpha S_j^\alpha + J^\beta S_i^\beta S_j^\beta + J^\gamma S_i^\gamma S_j^\gamma \nonumber \\
    &+ \Gamma(S^\alpha_iS^\beta_j + S^\beta_iS^\alpha) + D(S^\alpha_iS^\beta_j - S^\beta_iS^\alpha_j) \Big],\label{distortedJKG}
  \end{align}
  where the bond labeling convention remains the same and $D$ denotes the DM exchange (see the Supplementary Material for exchange expressions). Estimating the strengths of these new exchanges, we find $J^x \simeq 2$ meV, $J^y \simeq 2$ meV, $J^z \simeq 6$ meV, $\Gamma \simeq -2$ meV and $D \simeq 2$ meV. When we return to the ideal octahedra limit, $J^z \rightarrow J+K$ and $J^x, J^y \rightarrow J$.

  Using the above parameters to model the distortion effects, the magnetic ground state is found to be stripy. We next analyzed the effect of increasing $|D|$ and $|\Gamma|$ while keeping $J^x$, $J^y$ and $J^z$ fixed. Combining LT and CMC analyses, we find that the DM exchange tends to stabilize stripy order. Indeed, upon varying $|D|/J^z$ and $|\Gamma|/J^z$ independently in the interval $[-100,100]\times[-100,100]$, we find only stripy order. Therefore, we predict that the ground state of Ba$_3$IrTi$_2$O$_9$ harbors a stripy magnetic ordering pattern. 

\section{Discussion and Conclusion}
In summary, we have derived a general $j_\text{eff}=1/2$ spin model on a triangular lattice based on lattice symmetry considerations and the interplay between strong electron correlations, large SOC and a local octahedral crystal field environment. By employing Luttinger-Tisza and classical Monte Carlo, we have identified six distinct magnetic ground states supported by our model. We then determined the electronic band structure and nature of the states near the Fermi level in the layered triangular compound Ba$_3$IrTi$_2$O$_9$ from \textit{ab-initio} calculations, and found that it is well described by the pseudo-spin $j_\text{eff}=1/2$ picture. Applying our \textit{ab-initio} calculations to our general spin model, we predict that Ba$_3$IrTi$_2$O$_9$ hosts a stripy magnetic ground state near the Kitaev limit. Our prediction for stripy order as the magnetic ground state of Ba$_3$IrTi$_2$O$_9$ can be verified by neutron scattering. A magnetic peak at ordering wave vector \textbf{M} in the Brillouin zone would correspond to stripy order according to our analysis.

A previous experimental study of Ba$_3$IrTi$_2$O$_9$ \cite{Dey2012} measured zero-field-cooled and field-cooled magnetic susceptibility in the temperature range of 2-400 K, as well as heat capacity measurements from 0.35-295 K. No long range magnetic ordering was reported in this experimental study; however, X-ray powder diffraction of Ba$_3$IrTi$_2$O$_9$ at room temperature revealed large site sharing between Ir$^{4+}$ and Ti$^{4+}$ ions of $(37 \pm 10)\%$. It is possible that the site sharing between Ir$^{4+}$ and Ti$^{4+}$ ions could lead to a nonmagnetic dilution of the triangular layers. Studies of nonmagnetic dilution have been recently undertaken on the honeycomb iridates AIrO$_3$ (A = Li, Na)\cite{ManniTokiwaGenenwart2014, DasGuoRoychowdhury2015}, with the observation of possible spin-glass behavior. Further studies of nonmagnetic dilution on triangular lattices, which may enhance frustration and induce a correlated paramagnetic phase, can be important both in understanding material properties and in realizing novel phases. Based on our work, we expect that stripy magnetic ordering will be observed in a pure sample. 
  
\begin{acknowledgements}
We thank E. K. H Lee and Kyusung Hwang for fruitful discussions. This work was supported by the NSERC of Canada and the Center for Quantum Materials at the University of Toronto. Computations were mainly performed on the GPC supercomputer at the SciNet HPC Consortium. SciNet is funded by the Canada Foundation for Innovation under the auspices of Compute Canada; the Government of Ontario; Ontario Research Fund for Research Excellence; and the University of Toronto. HYK is grateful for the hospitality of KITP at Santa Barbara. 
\end{acknowledgements}

\appendix
\section{Derivation of $j_\text{eff}=1/2$ spin model without octahedra distortion}
We derive the spin model in the limit of ideal octahedra by performing a strong coupling expansion in the limit where $U, J_H \gg \lambda_{SO} \gg t$, with $U$ being the strength of the Coulomb interaction, $J_H$ is Hund's coupling, $\lambda_{SO}$ is the SOC strength and $t$ are the tight binding parameters. We start by considering an on-site Kanamori Hamiltonian,
\begin{equation}
  H_0 = \sum_i \left(\frac{U-3J_H}{2}(N_i-5)^2 - 2J_H S_i^2 - \frac{J_H}{2}L_i^2\right),
\end{equation}
where $i$ denotes the Ir$^{4+}$ site index, and $N_i, S_i$ and $L_i$ are the total number, spin and angular momentum operator on the $i^{th}$ site.

To derive an effective spin Hamiltonian, we consider electron hopping through major nearest-neighbor $d$-orbital hopping channels as small perturbations on top of $H_0$ within the framework of second order perturbation theory. The kinetic part of the Hamiltonian (along a $z$-bond for instance) takes the general form
\begin{equation}
  T_{ij}^z + (T_{ij}^z)^\dagger = \sum_{\sigma,\alpha,\beta}\left(d^\dagger_{i\alpha\sigma}\widetilde{T}^z_{\alpha\beta} d_{j\beta\sigma} + d^\dagger_{j\beta\sigma}(\widetilde{T}^z)^\dagger_{\alpha\beta} d_{i\alpha\sigma}\right), \nonumber
\end{equation}
where $d^\dagger_{i\alpha\sigma}$ creates a $d$-orbital electron ($\alpha,\beta = yz,zx,xy$) on site $i$ with spin $\sigma$, and $\widetilde{T}^z$ is the hopping matrix along the $z$-bond connecting the orbitals on the two nearest-neighbor sites. In the ideal octahedra limit, the hopping matrix is restricted by $C_2$ rotation symmetry along the bond connecting the two sites and $C_{2z}$ symmetry about the bond center to the form (in the $|yz\rangle, |zx\rangle, |xy\rangle$ basis):
\begin{equation}
\widetilde{T}^z =
\begin{pmatrix}
  t_1 & t_2 & 0 \\
  t_2 & t_1 & 0 \\
  0  & 0 & t_3  
\end{pmatrix}.
\end{equation}

The resulting perturbed Hamiltonian is projected onto the $j_\text{eff} = 1/2$ subspace since the SOC is large. In terms of the $j_\text{eff} = 1/2$ spin operators, we arrive at the effective Hamiltonian
\begin{equation}
  H_\text{ideal} = \sum_{\alpha\beta(\gamma)\in\langle ij \rangle} \left(J {\bf S}_i\cdot {\bf S}_j + K S_i^\gamma S_j^\gamma +
  \Gamma (S_i^\alpha S_j^\beta + S_i^\beta S_j^\alpha)\right),
\end{equation}
written in the main text, with
\begin{align}
  J &= \frac{4}{27}\left(\frac{6t_1(t_1+2t_3)}{U-3J_H}+\frac{2(t_1-t_3)^2}{U-J_H}+\frac{(2t_1+t_3)^2}{U+2J_H}\right) \\
  K &= \frac{8J_H}{9}\left(\frac{(t_1-t_3)^2-3t_2^2}{(U-3J_H)(U-J_H)}\right) \\
  \Gamma &= \frac{16J_H}{9}\left(\frac{t_2(t_1-t_3)}{(U-3J_H)(U-J_H)}\right).
\end{align}

\section{Expression for spin exchanges when including octahedra distortion effects}
As discussed in the main text, the distorted octahedra add two new hopping parameters $t_2$ and $t'_2$. In terms of the hopping matrix of the previous section, we have (in the $|yz\rangle, |zx\rangle, |xy\rangle$ basis):
\begin{equation}
\widetilde{T}_\text{distorted}^z =
\begin{pmatrix}
  t_1 & t_2 & 0 \\
  t'_2 & t_1 & 0 \\
  0  & 0 & t_3  
\end{pmatrix}.
\end{equation}
This non-symmetric hopping matrix introduces a Dzyaloshinksii-Moriya exchange in the effective Hamiltonian and also splits the Heisenberg and Kitaev exchanges found in the ideal limit. Along a $z$-bond (sites 1 and 2), the Hamiltonian takes the form,
\begin{align}
  H_\text{distorted}^z = J_x S^x_1S^x_2 + J_yS^y_1S^y_2 + J_zS^z_1S^z_2 + \\ D\left(S^x_1S^y_2 - S^y_1S^x_2\right) + \Gamma\left(S^x_1S^y_2 + S^y_1S^x_2\right), \nonumber
\end{align}
with the new exchange parameters given by,
\begin{align*}
J_x = \ \frac{4}{27} \bigg( & \frac{2 t_1^2-4 t_1 t_3-2 t_2^2+t_2 t^\prime_2+t^{\prime 2}_2+2 t_3^2}{U-J_H}+ \\ &\frac{3 \left(2 t_1^2+4 t_1 t_3+t_2 t'_2-t^{\prime 2}_2\right)}{U-3 J_H}- \\ &\frac{(-2 t_1+t_2-t'_2-t_3) (2 t_1+t_2-t'_2+t_3)}{U + 2 J_H}\bigg),
\end{align*}
\begin{align*}
  J_y = \ \frac{4}{27} \bigg( & \frac{2 t_1^2-4 t_1 t_3+t_2^2+t_2 t'_2-2 t^{\prime 2}_2+2 t_3^2}{U-J_H}+ \\ &\frac{3 \left(2 t_1^2+4 t_1 t_3-t_2^2+t_2 t'_2\right)}{U-3 J_H}- \\ &\frac{(-2 t_1+t_2-t'_2-t_3) (2 t_1+t_2-t'_2+t_3)}{U + 2 J_H} \bigg),
\end{align*}
\begin{align*}
  J_z = \ \frac{4}{27} \bigg( & \frac{-t_1^2+2 t_1 t_3+2 t_2^2+5 t_2 t'_2+2 t^{\prime 2}_2-t_3^2}{U-J_H}- \\ &\frac{3 \left(-3 t_1^2-2 t_1 t_3+3 t_2 t'_2-t_3^2\right)}{U-3 J_H} + \\ &\frac{4 t_1^2+4 t_1 t_3+t_2^2-2 t_2 t'_2+t^{\prime 2}_2+t_3^2}{U+2J_H}\bigg),
\end{align*}
\begin{align*}
  \Gamma = \ & \frac{8}{9}J_H \frac{(t_2+t_2')(t_1-t_3)}{(U-3J_H)(U-J_H)},
\end{align*}
\begin{align*}
  D = \ & -\frac{8}{27}(t_2-t_2')\bigg(\frac{3(t_1+t_3)}{U-3J_H} + \frac{t_1-t_3}{U-J_H} + \frac{2t_1+t_3}{U+2J_H}\bigg).
\end{align*}

For Ba$_3$IrTi$_2$O$_9$, we find the following tight-binding parameters from \textit{ab-initio} calculations: $t_1 = 7.4$ meV, $t_2 = -13$ meV, $t_2' = -32$ meV and $t_3 = -119$ meV. Taking on-site Coulomb repulsion to be $U = 2.0$ eV and fixing $J_H/U = 0.2$, we find $J^x \simeq 2$ meV, $J^y \simeq 2$ meV, $J^z \simeq 6$ meV, $\Gamma \simeq -2$ meV and $D \simeq 2$ meV. We return to the ideal octahedra limit when $t_2 = t_2' = -23.5$ meV. In this limit we find $J \simeq 2$ meV, $K \simeq 4$ meV and $\Gamma \simeq -2$ meV.

\section{Real Space Spin Representation of Phases}

\begin{figure*}
  \centering
  \includegraphics[width=1.0\textwidth]{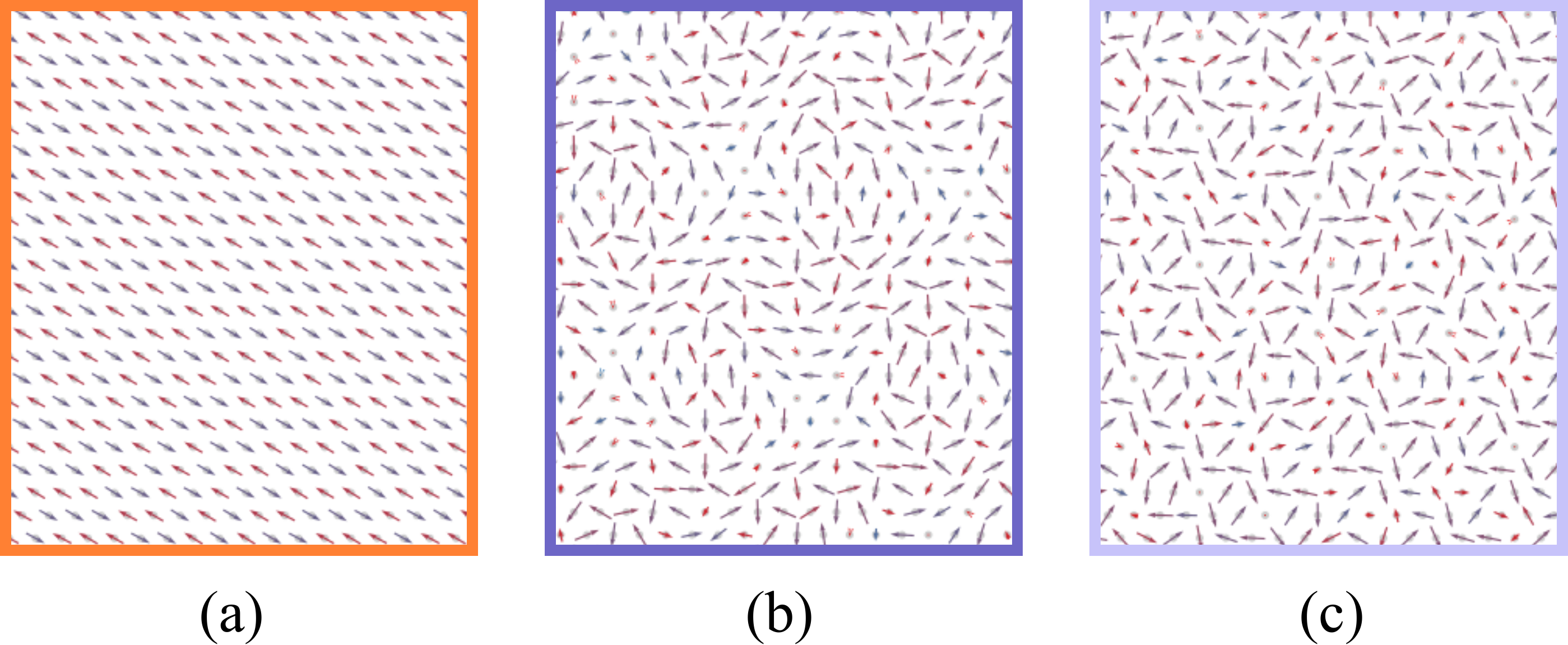}
  \caption{(Color online) Real space spin configuration examples of nematic (a), $\mathbb{Z}_2$ vortex crystal (b) and dual-$\mathbb{Z}_2$ vortex crystal phases. The orientation of each spin into (out of) the triangular plane is represented by the blueness (redness) of the color of the spins. The colored boundary around each phase corresponds to the color used to label the phase in Fig. \ref{fig:phasediagrams} of the main text.}
   \label{fig:realspacespins}
\end{figure*}

Real space spin representations of nematic, $\mathbb{Z}_2$ and dual-$\mathbb{Z}_2$ phases from CMC are shown in Fig. \ref{fig:realspacespins}. These phases are found near the boundary of the phase diagrams in Fig. \ref{fig:phasediagrams}, where $|\Gamma|$ is small. The nematic phase, Fig. \ref{fig:realspacespins}a, is found near the Kitaev limit, and characterized by uncorrelated antiferromagnetic chains that point along one of the three principal directions of the triangular lattice. The $\mathbb{Z}_2$ vortex crystal and dual-$\mathbb{Z}_2$ vortex crystal phases are shown in Figures \ref{fig:realspacespins}b and \ref{fig:realspacespins}c respectively, and are discussed in Refs. \onlinecite{roushochatzakis2012, becker2015}. The $\mathbb{Z}_2$ vortex crystal phase is incommensurate and features large vortices that span the lattice. It emerges upon perturbation of the Heisenberg model on the triangular lattice with a small Kitaev term. The dual-$\mathbb{Z}_2$ vortex crystal phase is related to the $\mathbb{Z}_2$ vortex crystal phase via Klein duality - a four-sublattice transformation of the spins that maps $J \rightarrow -J$ and $K \rightarrow 2J + K$. 

\bibliography{ba3irti2o9}

\end{document}